\documentclass[12pt]{iopart}
\usepackage{graphicx}
\usepackage{textcomp}
\usepackage{hyperref}
\usepackage{color}
\newcommand{\pss}{\psi_\ast} 

\newcommand{\ze}{\zeta} 
\newcommand{\zev}{\hat\zeta}
\newcommand{\vv}{\mathbf{v}} 
\newcommand{\cd}{\cdot}
\newcommand{\na}{\nabla} 
\newcommand{\ea}{e_a} 
\newcommand{\ma}{m_a}
\newcommand{\ee}{e_e} 
\newcommand{\me}{m_e}
\newcommand{\ei}{e_i} 
\newcommand{\mi}{m_i}
\newcommand{\Te}{T_e}
\newcommand{\Ti}{T_i}

\newcommand{\foe}{f_{1e}} 
\newcommand{\foi}{f_{1i}} 
\newcommand{\foa}{f_{1a}} 
\newcommand{\gef}{g_{e}}
\newcommand{\gif}{g_{i}}
\newcommand{\gaf}{g_{a}}
\newcommand{\fse}{f_{\ast e}} 
\newcommand{\fme}{f_{M e}}
\newcommand{\fmi}{f_{M i}} 
\newcommand{\fma}{f_{M a}}

\newcommand{\fze}{f_{0e}} 

\newcommand{\Foa}{F_{1a}}
\newcommand{\Fta}{F_{2a}}
\newcommand{\Foe}{F_{1e}}
\newcommand{\Fte}{F_{2e}}
\newcommand{\Foi}{F_{1i}}

\newcommand{\Oa}{\Omega_a} 

\newcommand{\p}{\partial} 
\newcommand{\Ev}{\mathbf{E}}
\newcommand{\Evz}{\mathbf{E}_0} 
\newcommand{\Evo}{\mathbf{E}_1}
 
\newcommand{\Bv}{\mathbf{B}}
\newcommand{\Rv}{\mathbf{R}}
\newcommand{\rv}{\mathbf{r}}
\newcommand{\Av}{\mathbf{A}_1} 
\newcommand{\Aps}{A_\psi}
\newcommand{\Ath}{A_\theta} 
\newcommand{\Aze}{A_\zeta}
\newcommand{\Apa}{A_\|} 
\newcommand{\nps}{\nabla\psi}
\newcommand{\nth}{\nabla\theta} 
\newcommand{\nze}{\nabla\zeta}
\newcommand{\bv}{\mathbf{b}} 
\newcommand{\kv}{\mathbf{k}} 

\newcommand{\Bvo}{\mathbf{B}_1} 
\newcommand{\vpa}{v_\|}

\newcommand{\kpa}{k_\|}
\newcommand{\kpe}{k_\perp}
\newcommand{\jpa}{j_\|}
\newcommand{\vpe}{v_\perp} 
\newcommand{\Epa}{E_\|}
 
\newcommand{\pho}{\phi_1}
\newcommand{\xie}{\xi_e}
\newcommand{\xii}{\xi_i}
\newcommand{\xia}{\xi_a}
\newcommand{\Ze}{Z(\xi_e)}
\newcommand{\Zi}{Z(\xi_i)}

\newcommand{\ose}{\omega_{\ast e}}
\newcommand{\osi}{\omega_{\ast i}}
\newcommand{\osa}{\omega_{\ast a}}
\newcommand{\ete}{\eta_e}
\newcommand{\eti}{\eta_i}
\newcommand{\xh}{\hat{x}}
\newcommand{\yh}{\hat{y}}
\newcommand{\zh}{\hat{z}}
\newcommand{\Bh}{\hat{B}}
\newcommand{\Bx}{B_x}
\newcommand{\ky}{k_y}

\begin{document}

\title[Current-driven electromagnetic]{A current driven electromagnetic
  mode in sheared and toroidal configurations}

\author{Istv\'{a}n Pusztai$^{1,2}$, Peter J Catto$^{1}$, 
Felix I Parra$^{1,3}$, Michael~Barnes$^{1,4}$}

\address{$^1$ Plasma Science and Fusion Center, Massachusetts
  Institute of Technology, Cambridge, MA 02139, USA} \address{$^2$
  Department of Applied Physics, Chalmers University of Technology and
  Euratom-VR Association, SE-41296 G\"oteborg, Sweden} \address{$^3$
  Department of Physics, University of Oxford, Oxford, OX1 3PU, UK}
\address{$^4$ Department of Physics, University of Texas at Austin,
  Austin, TX 78712, USA}

\ead{pusztai@chalmers.se}
\begin{abstract}
The induced electric field in a tokamak drives a parallel electron
current flow. In an inhomogeneous, finite beta plasma, when this
electron flow is comparable to the ion thermal speed, the Alfv\'{e}n
mode wave solutions of the electromagnetic gyrokinetic equation can
become nearly purely growing kink modes. Using the new "low-flow"
version of the gyrokinetic code {\sc gs2} developed for momentum
transport studies [Barnes et al 2013 {\it Phys. Rev. Lett.} {\bf 111}
  055005], we are able to model the effect of the induced parallel
electric field on the electron distribution to study the destabilizing
influence of current on stability. We identify high mode number kink
modes in {\sc gs2} simulations and make comparisons to analytical
theory in sheared magnetic geometry. We demonstrate reassuring
agreement with analytical results both in terms of parametric
dependences of mode frequencies and growth rates, and regarding the
radial mode structure.

\end{abstract}

\maketitle


\section{Introduction}

The radial gradient of electric current represents a source of free
energy in fusion plasmas which can drive or modify instabilities.  For
a sufficiently strong current gradient, kink modes can be
destabilized. The criterion for destabilization in a screw pinch was
derived in \cite{kadomtsev} with a magnetohydrodynamic formulation
(there referred to as the screw-instability) for high mode numbers.

Low mode number kink modes are important for the internal stability of
tokamaks. The $n=m=1$ internal kink mode is believed to be responsible
for the sawtooth instability \cite{sawtooth}. Most of the work done on
these modes uses a fluid formalism. However, accounting for kinetic
effects is important to reproduce all details of the evolution of such
instabilities. For instance, finite electron inertia can assist
collisionless reconnection that can modify the dynamics of $m=1$
internal kinks, as well as their coupling to ion sound waves, as
discussed in \cite{coppi}.  In addition, considering fluid ions and
kinetic electrons, collisional and diamagnetic effects on the $m=1$
mode were studied in \cite{drake}.  In spite of the recognized
importance of kinetic effects on kink modes, at present there are only
a limited number of numerical studies in the literature which employ
gyrokinetic \cite{cattolinear,frieman} simulations. The first is a
simulation of a sawtooth crash that was reported in \cite{hiroshi},
where a particle-in-cell (PIC) code neglecting ion finite Larmor
radius (FLR) effects was used to model the instability for straight
field lines. More recently, ideal-MHD internal kink and collisionless
$m=1$ tearing mode simulations were performed in a screw pinch
  geometry in \cite{mischenko} with the PIC code {\sc gygles}
\cite{gygles}. Gyrokinetic studies on modifications to kinetic
instabilities due to parallel current are also very limited. In
\cite{deng} the effects of equilibrium current on reversed shear
Alfv\'{e}n eigenmodes is studied using the PIC code {\sc gtc}
\cite{gtc}.  Moreover, the continuum gyrokinetic code {\sc gene}
\cite{gene} is used in \cite{pueschel} to study magnetic reconnection,
where alternating current sheets are modeled in a periodic slab
configuration. Linear gyrokinetic simulations of tearing modes in the
collisional–-collisionless transitional regime in a slab geometry are
presented in \cite{numata} using the A{\sc stro}GK code
\cite{agk}. These last two references introduce the parallel current
as a first-order gyrokinetic perturbation, rather than an unperturbed
drive term (part of the background distribution) as we do in this
article.

It is of interest to further develop our kinetic simulation capability
for current driven instabilities. Using the tools available in the new
version of the gyrokinetic code {\sc gs2} \cite{gs2}, developed for
intrinsic rotation studies in tokamaks \cite{barnesPRL}, we are now
able to model the destabilizing effect of the modifications to the
non-fluctuating electron distribution function due to an induced
electric field in a tokamak. In particular, current driven modes can
be studied using this continuum gyrokinetic code, as demonstrated
herein through simulations of high mode number kink modes with {\sc
  gs2}. The {\sc gs2} simulations presented here are radially
  local (flux tube), which inherently assumes a separation of the
  parallel and perpendicular scale lengths of perturbed
  quantities. Accordingly, in {\sc gs2}, only high mode number modes
  can be simulated, while global modes, such as the $n=m=1$ mode are
  beyond the region of applicability of local codes. The simulations
  are done in toroidal geometry and no simplifying assumptions (i.e.,
  regarding finite Larmor radius effects, kinetic treatment of
  different species, particle drifts etc.) are made to the
  Maxwell-gyrokinetic system apart from those consistent with the
  lowest order local gyrokinetic treatment. The new feature is the
  treatment of the modification to the electron distribution due to
  the induced electric field as an unperturbed drive term entering as
  a part of the non-fluctuating distribution. The code results will
be shown to be in very good agreement with the analytical calculations
we present.

The subsequent sections are organized as follows. First, in
Sec.~\ref{secGK} the electromagnetic gyrokinetic equations are derived
in toroidal geometry in the presence of an induced parallel electron
current. In Sec.~\ref{toroidal}, the dispersion relation of the high
mode number kink modes is derived in shearless toroidal geometry. The
effects of magnetic shear and the eigenmode structure are discussed in
Sec.~\ref{shearsec}. Finally, in Sec.~\ref{modechar} the analytical
results are compared to {\sc gs2} simulations, before we conclude in
Sec.~\ref{secconc}.


\section{Electromagnetic gyrokinetic equations with induced current}
\label{secGK} 
The induced electric field driven part of the non-fluctuating electron
distribution is similar to the solution of the Spitzer problem,
$C_l[f_{\rm Spitz}]=-(\ee/\Te)E_I \vpa f_{Me}$, where $C_l$ is the
linearized electron collision operator, $\fma=n_a[m_a/(2\pi
  T_a)]^{3/2}\exp[-m_a v^2/(2T_a)]$ is the Maxwell distribution, with
the density $n_a$, temperature $T_a$, mass $m_a$ and charge $\ea$ of
species $a$ (ions and electrons are denoted with the indices $a=i$ and
$e$, respectively).  Furthermore, $E_I$ denotes the induced parallel
electric field, $v^2=\vv\cd\vv$ and $\vpa=\vv\cd\bv$, with $\vv$ the
velocity and $\bv$ the unit vector in the direction of the equilibrium
magnetic field $\Bv_0$. The Spitzer function is proportional to
$\vpa$, but it may have a non-trivial speed dependence. However, as it
will be shown later through simulations, the exact velocity space
structure of $f_{\rm Spitz}$ is unimportant for the instability to be
investigated here. Therefore, the induced electric field effects will
be modeled simply by allowing for a parallel drift velocity.

To derive the linearized gyrokinetic equation it is convenient to use
the unperturbed total energy, $E=v^2/2+(\ea/\ma)\phi_0$, the
  magnetic moment, $\mu=\vpe^2/(2 B_0)$, and the canonical
angular momentum $\pss=\psi-(\ma c/\ea)R\zev \cd \vv$ as phase-space
variables. Here, $\vpe^2=v^2-\vpa^2$, $B_0=|\Bv_0|$, $\phi_0$ is
the non-fluctuating part of the electrostatic potential, $c$ denotes
the speed of light, $R$ is the major radius, $2 \pi\psi$ is the
poloidal magnetic flux, and $\zev=\na\ze/|\na\ze|$, with the toroidal
angle $\ze$ and $R|\na\ze|=1$. The unperturbed Vlasov operator
$d_t\doteq\p_t+\vv\cd\na+[(\ea/\ma) \Evz+\Oa\vv\times\bv]\cd\na_v$
acting on functions of only $E$ and $\pss$ vanishes in a toroidally
symmetric system which we shall consider. We have introduced $\Oa=\ea
B_0/(\ma c)$, with $\Evz=-\na\phi_0+E_I$. The time independent
piece of the distribution functions should be close to
\begin{equation}
f_{\ast a}(\pss,E)=\eta_{\ast a}\left(\frac{\ma}{2\pi T_{\ast
    a}}\right)^{3/2}\exp\left[-\frac{\ma E}{T_{\ast a}}\right],
\label{fsa}
\end{equation}
where $T_{\ast a}=T_a(\psi\rightarrow\pss)$, and the pseudo-density is
$\eta_{\ast a}=n_{\ast a} \exp[\ea\phi_{0\ast}/T_{\ast a}]$ with
$\phi_{0\ast}=\phi_0(\psi\rightarrow\pss)$ and $n_{\ast
  a}=n_a(\psi\rightarrow\pss)$. Note that $\phi_0$, $n_a$ and $T_a$ are
assumed to be flux functions.  We consider $\phi_0=0$. By
construction, $f_{\ast a}$ reduces to a Maxwellian as $\pss
\rightarrow \psi$.

In order to account for the electron flow due to the induced electric
field, we model the non-fluctuating electron distribution by
\begin{equation}
\fze=\fse(\pss,E)+f_s(\Rv_e,E,\mu),
\label{fzedef}
\end{equation}
where $f_s=-m_e\vpa u \fme/T_e$,
$\Rv_a=\rv+\Omega_a^{-1}\vv_\perp\times\bv$ is the particle guiding center,
and $\rv$ is the particle position. Furthermore, $\vv_\perp=\vv-\vpa
\bv$, and the parallel electron flow velocity is $-u$, where $u>0$ is
allowed to be comparable to the ion thermal speed
$v_i=(2T_i/m_i)^{1/2}$, and the sign of $u$ is chosen so that the unperturbed 
current density is $j_0=e n_e u$. For the electron flow to be divergence
free, $u\propto B_0$.   

The linearized kinetic equation for the fluctuating part of the
electron distribution $\foe$ can be written as
\begin{equation}
d_t \foe = -\frac{\ee}{\me}\left(\Evo+\frac{1}{c}\vv\times\Bvo
\right)\cd\na_v \fze,
\label{kine1}
\end{equation}
 where collisions are neglected since we are interested in the
   tokamak core, where the collision frequency is small. The
 fluctuating parts of the electric and magnetic fields are denoted by
 $\Evo$ and $\Bvo$,respectively. We note, that the induced
   electric field $E_I$ is accounted for by retaining its effect on
   the non-fluctuating distribution, i.e. keeping $f_s$ in $\fze$. The
   induced electric field is negligible in the $d_t$ term of
   (\ref{kine1}), since electron-ion drag requires it to be the same
   order as a collisional correction.
 
We represent the perturbed vector potential as
$\Av=\Aps\nps+\Ath\nth+\Aze\nze$, and work in the Coulomb gauge
($\na\cd\Av=0$).  Using  $\Evo=-\na\pho-c^{-1}\p_t\Av$ we obtain
\begin{eqnarray}
\left(\Evo+\frac{1}{c}\vv\times\Bvo \right)\cd\na_v
\fse=\label{gfse}\\ -\left[ \vv\cd\na\pho+\frac{\vv}{c}\cd\left(\nps
  \frac{\p \Aps}{\p t}+\nth\frac{\p \Ath}{\p t}
  +\nze\frac{\p \Aze}{\p t} \right)\right]\frac{\p \fse}{\p
  E}\nonumber \\ +\frac{\me}{\ee}\left[c\frac{\p\pho}{\p\ze}+\frac{d
    \Aze}{d t} -\vv\cd\left(\nps\frac{\p \Aps}{\p \zeta}+\nth\frac{\p
    \Ath} {\p \zeta}+\nze\frac{\p \Aze}{\p \zeta}
  \right)\right]\frac{\p \fse}{\p \pss} \nonumber ,
\end{eqnarray}
where we define $\p_t\Aze+\vv\cd\na \Aze=d_t \Aze$. At this point we may
neglect finite orbit width corrections to the kinetic equation by
replacing $\p_E\fse$ by $\p_E\fme$ and $\p_{\psi\ast}\fse$ by
$\p_\psi\fme$.

The preceding analysis for $\fse$ is essentially exact, however, we
simplify the analytic treatment for $f_s$ by considering large aspect
ratio $\epsilon=r/R\ll 1$ tokamak magnetic geometry with low
normalized pressure $\beta_i=8\pi p_i/B_0^2\ll 1$, where $r$ is the
minor radius and $p_i=n_i T_i$ is the ion pressure. We assume that the
gyrokinetic ordering is satisfied by any perturbed quantity $Q$,
namely $\bv\cd\na Q\ll |\na Q|$, $1/L \ll |\na \ln Q|$, and
$f_{1a}/f_{0a}\sim e_a \pho/T_a\ll 1$, where $L$ represents the
perpendicular scale length of background plasma
parameters. Additionally, we assume the $-\bv\cd\na\pho$ and the
$-\p_t\Apa/c=-\bv\cd\p_t\Av/c$ parts of $\Epa=\bv\cd\Evo$ to be
comparable in magnitude.

Next we consider $\na_v$ acting on $f_s$.  Neglecting the $\na_R \vpa$
term and the poloidal variation of $u$ as small in $\epsilon$, gives
\begin{equation}
\hspace{-10mm}\na_v f_s=-\frac{\vpa}{\Omega_e
  B_0}(I\Bv_0-R^2B_0^2\na\zeta)\frac{\p}{\p\psi} \left(\frac{\me u
  \fme}{\Te}\right)-\frac{\me u \fme}{\Te}\bv-\vv\frac{\me\vpa u
}{\Te}\frac{\p\fme}{\p E},
\label{navfs2}
\end{equation}
where $\Bv_0=I\na\zeta+\na\zeta\times\na\psi$ and we use
$\Bv_0\times\na\psi=I\Bv_0-R^2B_0^2\na\zeta$. We introduce the
  thermodynamic forces $\Foa=(\ln
  n_a)'+\left[m_av^2/(2T_a)-3/2\right](\ln T_a)'$ and $\Fta=(\ln n_a
  u)'+\left[m_av^2/(2T_a)-5/2\right](\ln T_a)'$, where we denote
  $\psi$-derivatives by ${}'$. Using
$\bv\cd(\Evo+c^{-1}\vv\times\Bvo)\approx\Epa-c^{-1}\vv_\perp\cd\na
\Apa$ we find
\begin{eqnarray}
\left(\Evo+\frac{1}{c}\vv\times\Bvo \right)\cd\na_v f_s=-\frac{\me
  u\vpa\fme}{\Te}\Fte\nonumber\\ \times\left\{\frac{I}{\Omega_e}
\left(\Epa-\frac{\vv_\perp}{c}\cd\na\Apa\right)
+\frac{\me}{\ee}\left[c\frac{\p\pho}{\p\ze}+\frac{d \Aze}{d t}
  \right.\right.\label{gfs} \\ \left.\left.-\vv\cd\left(\nps\frac{\p
    \Aps}{\p \zeta}+\nth\frac{\p \Ath} {\p \zeta}+\nze\frac{\p
    \Aze}{\p \zeta} \right)\right] \right\} \nonumber-\frac{\me u
  \fme}{\Te}\left(\Epa-\frac{\vv_\perp}{c}\cd\na\Apa\right)
\nonumber\\-\left(\frac{\me}{\Te}\right)^2\vpa u
\fme\vv\cd\left[\na\pho+\frac{1}{c}\left(\nps\frac{\p \Aps}{\p
    t}+\nth\frac{\p \Ath} {\p t}+\nze\frac{\p \Aze}{\p t}
  \right)\right]\nonumber
\end{eqnarray}

Defining
\begin{eqnarray}
g_e=\foe+\frac{\ee\pho}{\Te}\fme\left(1-\frac{\me u \vpa}{\Te}\right)+
\Aze\fme\left[\Foe- \frac{\me u \vpa}{\Te}\Fte\right],
\label{gedef}
\end{eqnarray}
and, combining (\ref{kine1}), (\ref{gfse}) and (\ref{gfs}), we derive
the kinetic equation governing this portion of the distribution
function
\begin{eqnarray}
\frac{d \gef}{dt}=-\frac{\ee}{\me}\left(1-\frac{\me u
  \vpa}{\Te}\right)\left[\frac{\p \pho}{\p
    t}-\frac{\vv}{c}\cd\left(\nps \frac{\p \Aps}{\p t}+\nth\frac{\p
    \Ath}{\p t} +\nze\frac{\p \Aze}{\p t} \right) \right]\frac{\p
  \fme}{\p E}\nonumber\\ -c\left[\frac{\p \pho}{\p
    \ze}-\frac{\vv}{c}\cd\left(\nps \frac{\p \Aps}{\p
    \ze}+\nth\frac{\p \Ath}{\p \ze} +\nze\frac{\p \Aze}{\p \ze}
  \right) \right]\label{geq}\\ \times\left[\Foe-\frac{\me u
    \vpa}{\Te}\Fte\right] +
\left(\Epa-\frac{\vv_\perp}{c}\cd\na\Apa\right)\left[Rc\frac{\me u
    \vpa}{\Te}\fme\Fte+\frac{\ee}{\Te}u\fme\right]\nonumber
\end{eqnarray}
To obtain this equation we made use of the fact that $d_t$ vanishes
when acting on $\p_E\fse$ and $\p_{\psi\ast}\fse$, and thus it
approximately vanishes when acting on $\p_E\fme$ and $\p_{\psi}\fme$,
as finite orbit width effects are neglected. Furthermore, we used
$\vv\cd\na\pho=(d_t-\p_t)\pho$.

Following a procedure similar to that in \cite{cattolinear} we can
derive the gyro-kinetic equation. After a transformation to gyro
center variables, a gyro-phase average of the kinetic equation
(\ref{geq}) is performed and finite orbit width effects are neglected
where appropriate. We neglect compressional magnetic perturbations as
small in the normalized pressure $\bv\cd\Bv_1/B_0\sim
\beta_i e \pho/T_i$.  For electrons we also neglect FLR
corrections. We note that $\vv_\perp\cd\na\Apa$ vanishes upon
gyro-phase averaging and the $\propto \Epa Rc$ term in the last line
of (\ref{geq}) is small in the gyrokinetic ordering and therefore can
be neglected. We arrive at the result
\begin{eqnarray}
\frac{\p \gef}{\p t}+(\vpa\bv+\vv_{de})\cd\na \gef =
\frac{\ee}{\Te}\left( \frac{\p \pho}{\p t}-\frac{\vpa}{c}\frac{\p
  \Apa}{\p t}\right)\left(1-\frac{\me }{\Te}u\vpa\right) \fme\nonumber
\\ - c \fme\left( \frac{\p \pho}{\p \zeta}-\frac{\vpa}{c}\frac{\p
  \Apa}{\p \zeta}\right) \left[\Foe-\frac{\me u \vpa}{\Te}\Fte\right]
+\frac{\ee}{\Te}\fme u \Epa,
\label{dkel}
\end{eqnarray}
where the electron drift velocity of the guiding center in the
equilibrium magnetic field is $\vv_{de}$, and the parallel component
of the fluctuating vector potential is
$\Apa=I(\Aze+\Ath/q)/(BR^2)$. This form of $\Apa$ follows from
assuming straight field line coordinates,
i.e. $\Bv_0\cd\na\theta=I/qR^2$ with $q$ a flux function. The full
fluctuating electron distribution $\foe$ and $\gef$ are related
by
\begin{eqnarray}
\gef=\foe+\frac{\ee \pho}{\Te}\fme\left(1-\frac{\me
}{\Te}u\vpa\right).
\label{foegefs}
\end{eqnarray}
where a term $\Aze\fme[\Foe-\me u \vpa \Fte/\Te]$
has been neglected as small in our ordering. The magnitude of this
term will be quantified in the beginning of Sec.~\ref{toroidal} when a
specific form for the perturbations will be assumed.

We keep FLR corrections when deriving the ion gyrokinetic equation to
obtain the usual result
\begin{eqnarray}
\frac{\p \gif}{\p t}+(\vpa\bv+\vv_{di})\cd\na \gif \nonumber \\=
\frac{\ei}{\Ti}\fmi\left( \frac{\p \langle\pho\rangle}{\p
  t}-\frac{\vpa}{c}\frac{\p \langle\Apa\rangle}{\p t}\right) - c \fmi
\left( \frac{\p \langle\pho\rangle}{\p \zeta}-\frac{\vpa}{c}\frac{\p
  \langle\Apa\rangle}{\p \zeta}\right)\Foi
 \label{gkion}, 
\end{eqnarray}
where $\langle\cd\rangle$ denotes a gyro-phase average at fixed
guiding center position, and the relation between $\gif$ and $\foi$ is
given by
\begin{equation}
\gif=\foi+\frac{\ei \pho}{\Ti}\fmi, 
\label{foigifs}
\end{equation}
where again, a term $\Aze\fmi\Foi$ has been neglected for our
ordering.

So far we have derived the linearized electromagnetic gyrokinetic
equations, where we allow for a parallel flow of electrons. We assumed
large aspect ratio and small beta, and neglected $\Ev_0$ and electron
FLR effects, but otherwise the equations (\ref{dkel}-\ref{foigifs}) are
still rather general. In the next section we shall derive a dispersion
relation for the high mode number kink modes, where further
approximations regarding the magnetic geometry and the mode structure
will be made.


\section{Dispersion relation of the high mode number kink mode}
\label{toroidal}
In this section we assume a flute like mode structure for the
perturbed quantities $\propto \exp(-i\omega t +im\theta-i n\zeta)$,
where the fluctuations are elongated along magnetic field lines with
$m\approx n q\gg 1$, where $q\sim 1$ is the safety factor. From
$\na\cd\Av=0$ we find $\Ath\approx\Aze r^2 n/(R^2 m)$ and thus
$\Apa\approx I\Aze[1+r^2/(qR)^2]/(BR^2)\approx\Aze/R$ and
$\bv\cd\na=iI(m-nq)/(qBR^2)\approx i\kpa$ with $\kpa$ the parallel
wave number. The size of $\Apa\sim \kpa c \pho/\omega$ is set by
assuming $E_\|\approx 0$.  We consider a pure plasma, but allow the
ion charge number $Z$ to be different from 1. By assuming $L$ to be
comparable with the minor radius of the device, we can neglect
magnetic drifts as small in $\epsilon$ as compared to the diamagnetic
drifts when deriving our dispersion relation. The justification of
neglecting magnetic drifts, which is a good approximation at long
wavelengths, will be further discussed towards the end of this
section.  Finally, the mode frequency $\omega$ is assumed to be
comparable or larger than the diamagnetic frequency $\sim
(n_icT_i/e_i) (\ln p_i)'$. We find that the $A_\zeta$ terms
neglected in the derivation of (\ref{foegefs}) and (\ref{foigifs}) are
smaller than the $\pho$ terms by $(\kpa
v_i/\omega)(\rho_i/L)(q/\epsilon)$, with $\rho_i=v_i/\Omega_i$ the ion
gyro radius.

From quasineutrality we have  
\begin{equation}
0=\sum_{a}\ea\int d^3v\,\left[\gaf+(\foa-\gaf)\right],
\label{qn1}
\end{equation}
where $\foa-\gaf$ for electrons and ions are given in (\ref{foegefs})
and (\ref{foigifs}). The velocity integrals of the $\foa-\gaf$ parts of
the distributions are straightforward to evaluate. Using
quasineutrality for the unperturbed densities, equation (\ref{qn1})
reduces to
\begin{equation}
\sum_{a}\ea\int d^3v \,\gaf =\frac{\ee^2
  n_e}{\Te}\left(1+\frac{Z\Te}{\Ti}\right)\pho,
\label{qn2}
\end{equation} 
where $Z=-\ei/\ee$ denotes the ion charge number.  The fluctuating
parallel current $\jpa$ is given by
\begin{equation}
\jpa=\sum_{a}\ea\int d^3v \,\vpa \left[\gaf+(\foa-\gaf)\right].
\label{jpa1}
\end{equation} 
Again, the velocity integrals of $\foa-\gaf$ can be readily evaluated
using the relations (\ref{foegefs}) and (\ref{foigifs}), to find
\begin{equation}
\sum_{a}\ea\int d^3v \, \vpa \gaf =\jpa-\ee n_e
u \frac{\ee}{\Te}\pho.
\label{jpa2}
\end{equation} 

Note that the velocity integrals of (\ref{qn1}-\ref{jpa2}) are taken
at fixed particle position, while the $g_a$ appearing in the
gyrokinetic equations are functions of the guiding center position.

When the magnetic drifts are neglected the gyrokinetic equations are
of the form
\begin{equation}
\p_t \gaf+ \vpa \bv\cd\na \gaf = {\rm RHS}_a,
\label{schem1}
\end{equation}
where ${\rm RHS}_a$ represents the right hand sides of (\ref{dkel})
and (\ref{gkion}) for $a=e$ and $i$, respectively. We integrate
(\ref{schem1}) over the velocity space at fixed particle position and
sum over species to find
\begin{equation}
\frac{\p}{\p t}\left(\sum_{a}\ea\int d^3v \,\gaf \right)+
\Bv\cd\na\left( \sum_{a}\frac{\ea}{B}\int d^3v\, \vpa \gaf
\right)=\sum_{a}\ea\int d^3v \, {\rm RHS}_a .
\label{schem2}
\end{equation}
The velocity integrals are to be performed in $E$ and $\mu$ variables,
and the Jacobian is $B/|\vpa|$, leading to the form of the second term of
(\ref{schem2}).  For electrons the integral of ${\rm RHS}_e$ gives
\begin{eqnarray}
\ee\int d^3v \, {\rm RHS}_e=\frac{\ee^2
  n_e}{T_e}\left(\frac{\p\pho}{\p t}+\frac{u}{c}\frac{\p \Apa}{\p
  t}\right)\nonumber \\-\ee n_e c \frac{\p\pho}{\p \zeta}\frac{\p\ln
  n_e}{\p \psi}-\ee n_e u \frac{\p \Apa}{\p \zeta}\frac{\p\ln (n_e
  u)}{\p \psi} +\frac{\ee^2}{\Te}n_e u \Epa
\label{intrhse1}
\end{eqnarray}
 For ions we need to account for FLR effects. We use $\int
 d^3v\,\fmi\langle\pho\rangle J_0(\kpe \vpe/\Omega_i)=\pho\int
 d^3v\,\fmi J_0^2(\kpe \vpe/\Omega_i)\approx n_i \pho
 [1-(\kpe\rho_i)^2/2]$ to evaluate the integral for the ions through
 first order in $\alpha_i=(\kpe\rho_i)^2/2$, where $\kpe$ is the
 perpendicular wave number. As a result we find
\begin{eqnarray}
\hspace{-8mm}\ei\int d^3v \, {\rm RHS}_i=(1-\alpha_i)\frac{\ei^2n_i}{\Ti}\frac{\p
  \pho}{\p t}- c \ei n_i\left[(1-\alpha_i)\frac{\p \ln p_i}{\p
    \psi}-\frac{\p \ln \Ti}{\p \psi}\right]\frac{\p \pho}{\p \zeta}.
\label{intrhsi1}
\end{eqnarray}
We then substitute (\ref{qn2}), (\ref{jpa2}), (\ref{intrhse1}) and
(\ref{intrhsi1}) into (\ref{schem2}) and use quasineutrality to find
\begin{eqnarray}
\Bv\cd\na\left(\frac{\jpa}{B}-\frac{\ee n_e
  u}{B}\frac{\ee}{\Te}\pho\right)=-\alpha_i\frac{\ee^2
  n_e}{T_e}\frac{Z \Te}{\Ti}\frac{\p \pho}{\p t} \nonumber
\\ +\frac{\ee^2 n_e}{T_e}\frac{u}{c}\frac{\p \Apa}{\p t}+\alpha_i\ei
n_i c \frac{\p \ln p_i}{\p \psi}\frac{\p\pho}{\p \zeta}-\ee n_e u
\frac{\p \Apa}{\p \zeta} \frac{\p \ln (n_e u)}{\p \psi} + \frac{\ee^2
  n_e}{\Te}u\Epa.
\label{eqjpa01}
\end{eqnarray}
Realizing that the $\bv\cd\na\pho$ term on the left hand side of
(\ref{eqjpa01}) together with the $\p_t \Apa$ term on the right hand
side of (\ref{eqjpa01}) exactly cancel with the $\Epa$ term, we can
simplify  to obtain
\begin{equation}
i\kpa\jpa=i\omega\tau\alpha_i\frac{\ee^2 n_e}{T_e}\pho-i n c \ei n_i
\alpha_i\pho\frac{\p \ln p_i}{\p \psi}+i n u \ee n_e \Apa \frac{\p \ln
  (n_e u)}{\p \psi},
\label{eqjpa02}
\end{equation}
where we employ the mode structure $\exp(-i\omega
t+im\theta-in\zeta)$ and define $\tau=Z\Te/\Ti$. Introducing the
background current gradient and the pressure gradient driven diamagnetic
frequencies
\begin{equation}
\omega_{\ast e}^j=\frac{nc\Te}{\ee}\frac{\p\ln j_0}{\p
  \psi},\qquad\omega_{\ast i}^p=\frac{nc\Ti}{\ei}\frac{\p\ln p_i}{\p
  \psi},
\end{equation}
with $j_0=e n_e u$ and $p_i=n_i T_i$, and multiplying
(\ref{eqjpa02}) by $-i\kpa c T_e/(\ee^2n_e)$ we find
\begin{equation}
\frac{\kpa^2c\Te}{\ee^2 n_e}\jpa=\alpha_i\tau\kpa
c\pho(\omega-\omega_{\ast i}^p)+\omega_{\ast e}^j\kpa u \Apa.
\label{eqjpa03}
\end{equation}
Then we employ the parallel Amp\`{e}re's law
$\kpe^2\Apa=(4\pi/c)\jpa$, and recall $\beta_i=8\pi
p_i/B^2=(v_i/v_A)^2$, where $v_A=[B^2/(4\pi m_i n_i)]^{1/2}$ is
the Alfv\'{e}n speed, to rewrite the left hand side of (\ref{eqjpa03})
as $\tau (v_i\kpa)^2\alpha_i\Apa/\beta_i$. We focus on the
$\Epa\approx 0$ limit, that is $\pho\approx \omega \Apa /(\kpa
c)$. This approximation will be justified at the end of this
section. We use this relation to eliminate $\pho$ from (\ref{eqjpa03})
in favor of $\Apa$, and then divide by $\tau(\kpe\rho_i)^2/2$ to
obtain the dispersion relation
\begin{equation}
\omega(\omega-\omega_{\ast i}^p)=
\frac{(v_i\kpa)^2}{\beta_i}-\frac{\omega_{\ast e}^j\kpa u}{\tau\alpha_i}.
\label{predisp}
\end{equation}
The solution of (\ref{predisp}) for the mode frequency is then
\begin{equation}
\omega=\frac{\omega_{\ast i}^p}{2}\pm\left[\left(\frac{\omega_{\ast
      i}^p}{2}\right)^2+\frac{(v_i\kpa)^2}{\beta_i}-\frac{2\omega_{\ast
      e}^j\kpa u}{\tau(\kpe\rho_i)^2}\right]^{1/2}.
\label{disp0}
\end{equation}
This result is consistent with Equation (15) of \cite{sperling}, which
was derived in a shearless slab geometry.  In the $\omega_{\ast
  i}^p\ll\omega$ limit (\ref{disp0}) reduces to
\begin{equation}
\omega=\pm\left(\frac{(v_i\kpa)^2}{\beta_i}-\frac{2\omega_{\ast
    e}^j\kpa u}{\tau(\kpe\rho_i)^2}\right)^{1/2}.
\label{disp1}
\end{equation}

For a given wave number if the electron flow speed $u$ or the
normalized pressure $\beta_i$ is sufficiently small the first term
dominates on the right hand side of (\ref{disp1}), and the solution is
an Alfv\'{e}n wave with purely real frequency $\omega^2\approx(v_A
\kpa)^2$. However, for high enough $\beta_i$, $u$ and $\omega_{\ast
  e}^j$ the second term might exceed the first and, depending on the
relative sign of $\kpa$ and u, (\ref{disp1}) describes either a pair
of stable modes with purely real frequencies or a purely growing and a
purely damped mode.

For the rest of this section we will be concerned with the purely
growing mode driven by the current gradient. Clearly, decreasing the
perpendicular wave number of the mode increases the growth rate of the
mode. Since the first term in (\ref{disp1}) is quadratic and the
second term is linear in $\kpa$, there is an optimal value of the
parallel wave number, $k_{\| o}$, where the mode has the highest
growth rate, $\gamma$. When the plasma parameters and the perpendicular
wave number are fixed the optimum is
\begin{equation}
k_{\| o}=\frac{u \beta_i \omega_{\ast e}^j}{ v_i^2(\kpe\rho_i)^2 \tau},
\label{optimalkpa}
\end{equation}
and the growth rate corresponding to $k_{\| o}$ is
\begin{equation}
\gamma_o=\frac{u\sqrt{\beta_i}\omega_{\ast e}^j }{v_i (\kpe\rho_i)^2 \tau}.
\label{growthrate}
\end{equation}
When $\omega_{\ast e}^j\sim\omega_{\ast i}^p$ and $\tau\sim 1$, the
assumption $\omega_{\ast i}^p\ll |\omega|$ used to obtain
(\ref{disp1}) is satisfied if $1\ll u\sqrt{\beta_i}/(v_i\alpha_i)$. As
long as there is a finite plasma beta and electron current, one can always
find sufficiently small perpendicular wave number for which this
relation is satisfied in the $\rho_i/L\rightarrow 0$ limit.  In this
case, neglecting magnetic drifts in the gyrokinetic equation is also
justified as long as the pressure length scale is much smaller than
the major radius.

It is shown at the end of \ref{appqn}, the perturbed quasineutrality
equation can be written in the form $0=[\pho-\omega \Apa/(\kpa
  c)]G_1+G_2$, where $G_1$ is a dimensionless function of order unity
(as long as $\omega/(\kpa v_e)$ is not too large) and $G_2$ is small
in $\alpha_i$. Thus, neglecting the small correction from $G_2$, the
approximate quasineutrality equation $0=[\pho-\omega \Apa/(\kpa
  c)]G_1$ is satisfied either if $G_1=0$, or $\pho-\omega \Apa/(\kpa
c)=0$, that is if $\Epa=0$, which we assumed in deriving
(\ref{disp1}). The case $G_1=0$ includes drift wave solutions and the
strongly damped modes corresponding to electrostatic roots of the
uniform plasma dispersion relation in the presence of electron flow.


\section{Magnetic shear effects}
\label{shearsec}

To obtain simple analytical results in Section \ref{toroidal} we
neglected magnetic drifts and assumed a flute like mode structure
(with no radial variation). The mode tends to be more unstable at low
perpendicular wave numbers, thus it is appropriate to neglect the
magnetic drifts, $\vv_{di}\cd\kv_\perp\ll\omega$.

Due to the preceding assumptions, the result (\ref{disp0}) is
formally the same as what one would obtain solving the problem in a
shearless slab geometry \cite{sperling}. The only difference between
a torus and a slab is that $ \omega_{\ast e}\propto n$ and
$\kpe\rho_i\propto n$ have lower limits set by the lowest finite
toroidal wave number $n=1$. We note that in a shearless slab there is
no such periodicity constraint, and $\kpe\rho_i$ can get arbitrarily
small (thus $\gamma$ arbitrarily large) for sufficiently large
perpendicular wave lengths. This unphysical behavior is partly
resolved by taking finite magnetic shear into account, which is needed
for the magnetic geometry to be consistent with a substantial parallel
current. In this section we will study the consequences of a magnetic
shear in slab geometry.

We choose a coordinate system $\{\xh,\yh,\zh\}$ such that plasma
parameters vary in the $\xh$ direction, and consider a mode which is
sinusoidally varying in the $\yh$ direction with a corresponding wave
number $k_y$, while the magnetic field has the form $\Bv=B(\zh+\yh
x/L_s)$. The magnetic shear produces an $x$ variation in $\kpa$,
namely $\kpa(x)=\kpa(0)+k_y x/L_s$, and we choose the origin so that
$\kpa(0)=0$. We assume that the radial variations of the perturbed
quantities are faster than those of the unperturbed ones and
$\beta_i\ll 1$, thus the $\yh$ component of the electron flow can be
neglected ($\mathbf{u}=u \zh$) together with any change in the
magnitude of $\zh\cd\Bv$.

To obtain a dispersion relation in a sheared geometry we start with
(\ref{eqjpa03}) and insert parallel Amp\'{e}re's law together with
$\alpha_i\rightarrow \rho_i^2(k_y^2-\partial_{xx}^2)/2$ to find
\begin{equation}
\tau
\frac{v_i^2\kpa^2}{\beta_i}\frac{\rho_i^2}{2}(k_y^2-\p_{xx}^2)\Apa=\tau\kpa
c(\omega-\omega_{\ast
  i}^p)\frac{\rho_i^2}{2}(k_y^2-\p_{xx}^2)\pho+\omega_{\ast e}^j\kpa u
\Apa.
\label{eqjpa03b}
\end{equation}
Then, we assume $\Epa\approx 0$ to replace $\pho$ in (\ref{eqjpa03b})
by $\omega \Apa/(\kpa c)$, which is consistent with neglecting
$\mathcal{O}(\alpha_i)$ terms in the quasineutrality equation. Taking
the $y$-derivative of (\ref{eqjpa03b}) leads to the dispersion
relation in terms of the $x$-component of the perturbed magnetic field,
$\Bx$
\begin{equation}
\hspace{-1.2cm}-\omega(\omega-\omega_{\ast
  i}^p)\frac{\rho_i^2}{2}\left(\ky^2-\p_{xx}^2\right)
\left(\frac{\Bh}{\kpa}\right)+\frac{(\kpa
  v_i)^2}{\beta_i}\frac{1}{\kpa}\frac{\rho_i^2}{2}
\left(\ky^2-\p_{xx}^2\right)\Bh -\frac{\kpa u \omega_{\ast
    e}^j}{\tau}\frac{\Bh}{\kpa}=0,
\label{sde1}
\end{equation}
where $\Bh$ is defined by $\Bx=\Bh(x)\exp(-i\omega t+i \ky
y)$. The dispersion relation is essentially the same as in shearless
geometry, except for the linear $x$-dependence of $\kpa$, and that the
replacement $\p_x\rightarrow ik_x$ cannot be made. 

Recalling $\kpa=\ky x/L_s$ and introducing the dimensionless
``radial'' coordinate $X=\ky x$, (\ref{sde1}) can be rewritten in the
form
\begin{equation}
X\left(\p_{XX}^2-1\right)\Bh-\lambda\left(\p_{XX}^2-1\right)
\left(\Bh/X \right)-\sigma \Bh=0,
\label{eqH}
\end{equation}
where $\lambda=\omega(\omega-\omega_{\ast
  i}^p)L_s^2\beta_i/v_i^2\approx\omega^2 L_s^2\beta_i/v_i^2$ and
$\sigma=-2L_s\beta_i u\omega_{\ast e}^j/(\tau k_y^2 \rho_i^2
v_i^2)$. The boundary conditions for this eigenvalue problem in
$\lambda$ are given by the requirement that $\Bh(|X|\rightarrow
\infty)\rightarrow 0$. In (\ref{eqH}) $\sigma$ represents the drive
and $\lambda<-\beta_i[\omega_{\ast i}^p L_s/(2v_i)]^2$ corresponds to
an instability ${\rm Im}(\omega)>0$. During the analysis of the radial
eigenmodes we shall neglect $(\omega_{\ast i}^{p}/\omega)^2\ll 1$
corrections, and refer to the $\lambda<0$ solutions as unstable and
the $\lambda=0$ solutions as marginally stable modes. We will retain
$\omega_{\ast i}^p$ corrections in Section \ref{modechar}. We note
that reversing the sign of $\sigma$, that is, the relative sign of $u$
and $\kpa$, leads to the same eigenvalues, and the corresponding
eigenfunctions satisfy $\Bh(X)|_{-\sigma}=\Bh(-X)|_{\sigma}$. Thus,
henceforth we will analyze solutions corresponding to $\sigma>0$,
without loss of generality.

For small values of $X$, (\ref{eqH}) is dominated by the $\lambda$
term, that is solved by $\Bh=c_1 X\exp(X)+c_2 X\exp(-X)$. Accordingly,
the solutions are either linear or quadratic in $X$ around $X=0$. For
high values of $X$ the first term dominates (\ref{eqH}), leading to
the an exponential asymptotic behavior $\Bh(X)\propto\exp(\pm X)$,
consistent with the boundary conditions. To solve numerically we
rewrite the eigenvalue problem (\ref{eqH}) for $F=\Bh(X)/X$ as
\begin{equation}
X(XF''+2F'-XF)-\lambda(F''-F)-\sigma X F =0,
\label{eqF}
\end{equation}
and discretize it using a second order finite difference scheme. We
set $(XF)'-XF=0$ as the negative $X$, and $(XF)'+XF=0$ for the
positive $X$, boundary conditions to select solutions with the appropriate
asymptotic behavior. Then we numerically search for the eigenvalues
$\lambda$ and eigenfunctions $F$ of the system for a given $\sigma$.

\begin{figure}[htbp]
\includegraphics[width=1\textwidth]{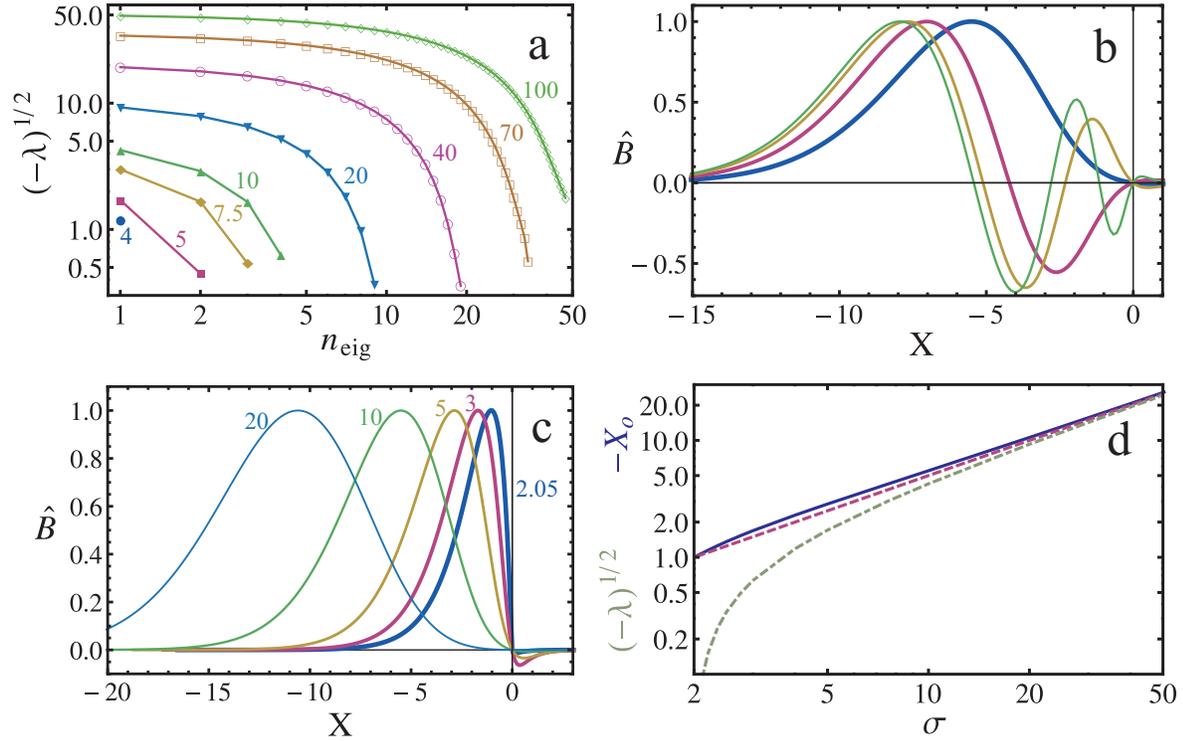}
\caption{Solutions of the eigenvalue problem (\ref{eqH}). (a)
  Normalized growth rates $(-\lambda)^{1/2}$ of unstable modes
  corresponding to $\sigma=\{4,\,5,\,7.5,\,10,\,20,\,40,\,70,\,100\}$.
  The number of unstable modes $\max(n_{\rm eig})$ increases with
  $\sigma$. (b) Radial eigenmodes $B(X)$ of the four unstable mode at
  $\sigma=10$. Lower growth rates correspond to more oscillatory
  structure. (c) The most unstable radial eigenmodes for
  $\sigma=\{2.05,\,3,5,\,10,\,20\}$. The distance of the location of the maximum
  of $|\Bh(X)|$, from $X=0$ increases with $\sigma$. (d) Solid curve:
  $-X_o$, where $X_o$ is the location of the maximum of $|\Bh(X)|$ for
  the most unstable mode at a given value of $\sigma$. Dash-dotted
  curve: normalized growth rate $(-\lambda)^{1/2}$ of the most
  unstable mode. Dashed curve: $\sigma/2$. }
\label{eigs}
\end{figure}

Figure \ref{eigs} shows solutions of the eigenvalue problem
(\ref{eqH}). We find that as the drive $\sigma$ is increased, more and
more unstable eigenfunctions appear, as illustrated in Figure
\ref{eigs}a showing the normalized growth rates $(-\lambda)^{1/2}$ of
all the unstable eigenmodes for different values of $\sigma$. On the
x-axis of Figure \ref{eigs}a, $n_{eig}$ denotes the ordinal number of
the unstable modes, with $n_{eig}=1$ corresponding to the most
unstable mode for each value of $\sigma$. In fact, a new unstable mode
appears as $\sigma$ exceeds $2N$ for every positive integer $N$. In
particular, no unstable mode exists for $|\sigma|\le 2$. Note that in
Figure \ref{eigs} the marginally stable ($\lambda=0$) modes for even
values of $\sigma$ are not shown.

For $\sigma=2N$, the marginally stable ($\lambda=0$) solutions of
(\ref{eqH}) are of the form $\Bh(X)=X \exp(-X)
{}_1F_1(1+\sigma/2,2;2X)=X \exp(X)P_\sigma(X)$ for $X\le 0$ and
$\Bh(X)=0$ for $X>0$. Here, ${}_1F_1$ denotes the Kummer confluent
hypergeometric function, and $P_\sigma$ is a polynomial with only
positive coefficients ($P_2=1$, $P_4=1+X$, $P_6=1+2X+2X^2/3$,
\dots). The derivative of the marginally stable solutions is
discontinuous at $X=0$, however it is resolved by a boundary layer at
$+0$ for $\sigma=2N+\delta$ with an arbitrarily small $\delta>0$ and a
corresponding small eigenvalue $\lambda$. The boundary layer connects
the $X\le 0$ solution vanishing at $X=0$, to a solution
$\propto\exp(-X)[1+2X\exp(2X){\rm Ei}(-2X)]$ for $X>0$, which is
finite at $X=0$. Here, ${\rm Ei}(x)=\mathcal{P}\int_{-\infty}^x\exp
(t)/t\,dt$ (for real values of $x$) denotes the Exponential integral,
where $\mathcal{P}$ indicates that the principal value is to be used
for $x\ge 0$. No marginally stable solution to (\ref{eqH}) exists if
$|\sigma|\ne 2N$, since in this case $X \exp(-X)
{}_1F_1(1+\sigma/2,2,2X)$ becomes divergent at $X\rightarrow -\infty$,
and the $\Bh(X\rightarrow -\infty)=0$ boundary condition cannot be
met.

When more than a single unstable eigenmode exists ($\sigma>4$), the
ones with lower growth rates exhibit a more oscillatory radial
structure, as illustrated in \ref{eigs}b showing the four unstable
modes for $\sigma=10$. In particular, the most unstable mode
(corresponding to the thickest curve in \ref{eigs}b) does not change
sign in the region $X<0$, while all the other unstable modes do.  This
is consistent with the behavior of the marginally stable modes, since
for increasing $N$ the number of roots of $P_\sigma(X)$ increases.

The amplitude $|\Bh|$ of the most unstable eigenmode has a maximum
close to the radial location where $\kpa(X)$ would maximize the local
dispersion relation (\ref{disp1}), that is $\kpa(X)\approx k_{|| o}$
with the optimal wave number $k_{|| o}$ given in
(\ref{optimalkpa}). 

In terms of $X$, the location of $\kpa(X)=k_{|| o}$ scales as
$X_o=-\sigma/2$ according to the local theory. As shown in
\ref{eigs}d, the location of the maximum amplitude (solid curve,
representing $-X_o$) follows this expectation (dashed curve,
$\sigma/2$) quite well. In the strongly driven ($\sigma\gg 1$) limit
the normalized growth rate $(-\lambda)^{1/2}$ of the most stable
eigenmode (dash-dotted curve in \ref{eigs}d) approaches the optimal
value, $\gamma_o$ given by (\ref{growthrate}). This value corresponds to
$(-\lambda)^{1/2}\rightarrow \sigma/2$. However, $\sigma=2$ gives
$(-\lambda)^{1/2}=0$ (that is, no unstable mode) in the sheared slab
model, while the local theory would predict a finite growth rate
equivalent with $(-\lambda)^{1/2}=1$. 

In conclusion, considering magnetic shear sets a stability limit in
terms of the drive at $\sigma=2$ in contrast to the shearless model
that predicts instability when $\beta_i$, the current gradient and the
flow speed $u$ are finite. In the shearless case the mode is always
allowed to pick the optimal parallel wave number.

The stability criterion of the mode $|\sigma|<2$ is equivalent to that
of the high mode number kink modes. Using the relations $u=j_0/(en_e)$,
$L_s=qR/s$, $s=(r/q)dq/dr$, $k_y=nq/r$ and $n=m/q$, with $m$ the
poloidal mode number, together with the definitions of $\sigma$ and
$\omega_{\ast e}^j$, one can rewrite the stability criterion
$|\sigma|<2$ as
\begin{equation}
\frac{4\pi r}{c B_\theta}\left|\frac{d j_0}{dr}\right|< 2m
\left|\frac{q'}{q}\right|,
\label{kado}
\end{equation}
as obtained from the magnetohydrodynamic energy principle in
\cite{kadomtsev} -- see Equation (2.29) therein. 



\section{Mode characteristics in toroidal geometry}
\label{modechar}

In this section the high mode number kink mode investigated in
Sections~\ref{toroidal} and \ref{shearsec} is studied numerically
using the gyrokinetic code {\sc gs2}. {\sc gs2} is free from the
simplifying assumptions made in Section \ref{shearsec}, except for the
radial locality and the scale separation $\kpa\ll\kpe$. In the
low-flow version of {\sc gs2} extra terms related to neoclassical
corrections to the non-fluctuating part of the distribution function
and the electrostatic potential are implemented for momentum transport
studies, as discussed in \cite{barnesPRL}. These quantities are
specified as inputs, normally calculated by the neoclassical code {\sc
  neo} \cite{neo}. This infrastructure can in principle be used to include any
modification to the non-fluctuating part of the distribution over a
velocity range of a few thermal speeds. We use it to include $f_s$, as
defined after (\ref{fzedef}), or more sophisticated Spitzer functions,
to study the effect of the induced electric field on
instabilities. Normally, we include only a parallel flow in {\sc gs2}
simulations, instead of a full Spitzer function since the results are
insensitive to the detailed form.
 
First we consider the parametric dependences of the mode frequency and
the growth rate, and compare {\sc gs2} simulations to predictions of the
sheared slab model (SSM) (\ref{sde1}).  The SSM results are obtained
by choosing the most unstable eigenmode from the numerical solution of
(\ref{eqF}). We use a 200 point radial grid, the extent of which is
adapted to the expected width of the eigenfunctions depending on the
value of $\sigma$.

The scans are performed about the following set of base-line
parameters: $u/v_i=1$, $\beta_i=0.01$, $a/L_u=3$,
$a/L_{Ti}=a/L_{Te}=a/L_n=0$, $k_y\rho_i=0.15$, $a/R=0.1$, $r/a=0.5$,
$s=1$, and $q=10$, where $d \ln u/dr=-1/L_u$, $d \ln n_e/dr=-1/L_n$,
$d \ln T_e/dr=-1/L_{Te}$, and $d \ln T_i/dr=-1/L_{Ti}$. We set the
density and temperature gradients to zero to avoid the appearance of
the usual gradient driven modes (otherwise, for this $\beta_i$,
magnetic shear and $L_n\sim a\sim L_T$ kinetic ballooning modes appear
and pollute the results, as in \cite{kinkEPS}). Then the only instability
drive is due to the gradient of the flow speed. The radial gradient of
the flow speed in the Ohmic current is due to density and electron
temperature gradients, thus our settings are not physically
consistent. However, by artificially choosing the parameters we obtain
a cleaner comparison between theory and simulations.

The binormal wave number and the aspect ratio are chosen to be small
so that magnetic drifts are not expected to affect the results
significantly. For a typical {\sc gs2} simulation only an extended
poloidal angle range of $\theta=(-\pi,\pi)$ is kept and $80$ 
grid points along the field line are used, since the eigenfunctions of
strongly driven modes are highly oscillatory and very localized in
$\theta$. The simulations use 20 untrapped pitch angle- and 14 energy
grid points. We neglect collisions and compressional magnetic
perturbations.

\begin{figure}[htbp]
\begin{center}
\includegraphics[width=1\textwidth]{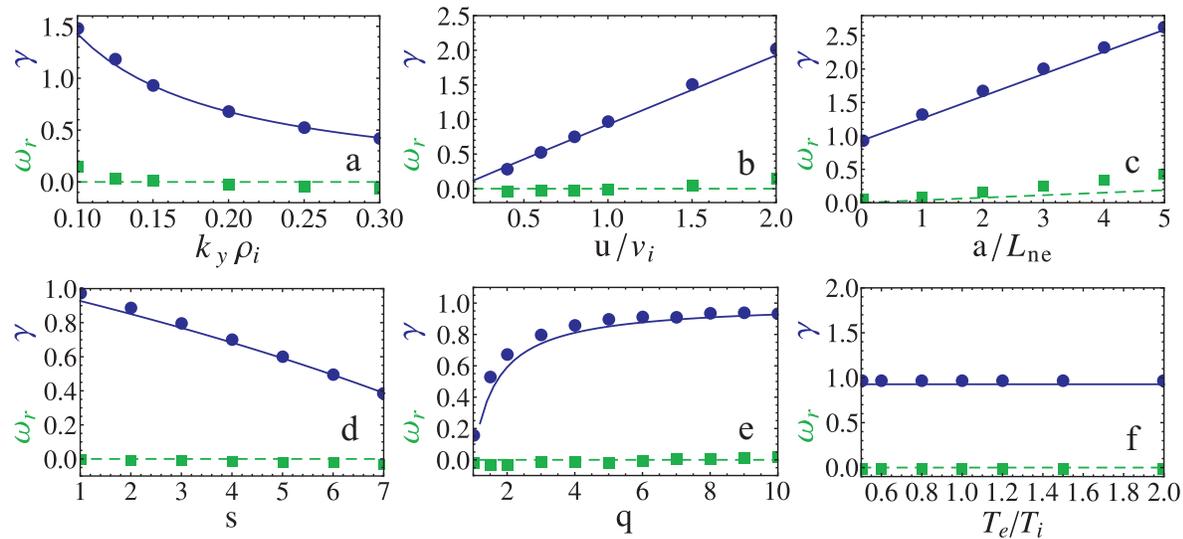}
\end{center} 
\caption{Parametric scalings of the growth rate $\gamma$ (solid line
  and circle markers) and real frequency $\omega_r$ (dashed lines and
  square markers) of the high mode number kink mode (given in $v_i/a$
  units). Markers represent {\sc gs2} simulations and lines are
  results of the sheared slab model. The figures depict the dependence
  on the following: (a) binormal wave number $k_y\rho_i$, (b) electron
  flow velocity $u/v_i$, (c) density gradient $a/L_n$, (d) magnetic
  shear $s$, (e) safety factor $q$, and (f) temperature ratio
    $T_e/T_i$. }
\label{compar}
\end{figure}

Figure \ref{compar} shows various parameter scalings around the
baseline parameter set. In a strongly driven situation ($|\sigma| \gg
2$), the growth rate is expected to be close to (\ref{growthrate}),
which helps in interpreting the numerical results. Since $\omega_{\ast
  e}^j\propto k_y$, and $\kpe\sim k_y$, we expect a $1/k_y$ dependence
of the growth rate, which is observed in Fig.~\ref{compar}a. Magnetic
drifts should be more important towards higher wave numbers. The good
agreement remains between {\sc gs2} and the SSM even at
$k_y\rho_i=0.3$ due to the very large aspect ratio $R/r=20$. The
growth rate is expected to increase linearly with the flow speed and
the mode should be stable at $u=0$ and this behavior is seen in
Fig.~\ref{compar}b. Similarly, the growth rate should exhibit the
linear dependence on $a/L_n$ as shown, where the mode is unstable at
$a/L_n=0$ due to the finite gradient in the flow speed, see
Fig.~\ref{compar}c.

To translate magnetic geometry parameters from the toroidal geometry
of {\sc gs2} to a sheared slab we use $1/L_s=s/qR$. Although the local
model can be used to explain certain parametric dependences of the
mode, it cannot provide predictions for the $L_s$ dependence. However
we know that as $\sigma\propto L_s$ drops below $2$ due to a
decreasing $L_s$, the mode should be completely stabilized. Thus we
expect increasing $s$ should reduce the growth rates, as seen in
Fig.~\ref{compar}d. Clearly, $q$ should have the opposite effect as
$s$, since $L_s\propto q/s$. Indeed, Fig.~\ref{compar}e shows
that the mode is stabilized with decreasing $q$. Also, when the mode
is strongly driven, $|\sigma|\gg 2$, the growth rate should become
independent of $L_s$, since the mode approaches the local
result. Hence, there is a saturation in the $q$-dependence of $\gamma$
towards higher values of $q$. When $\beta_i$ and $k_y\rho_i$ are
  held fixed the growth rate given in (\ref{growthrate}) normalized to
  $v_i/a$ is independent of $T_e/T_i$. The insensitivity of the result
  to the temperature ratio is demonstrated in Fig.~\ref{compar}d.

The real part of the frequency $\omega_r$ is proportional to the ion
diamagnetic frequency $\omega_{\ast i}^p$, which should be zero in
almost all the scalings of Fig.~\ref{compar}, since the ion pressure
gradient is zero. The only exception is the density gradient scaling,
Fig.~\ref{compar}c, where $\omega_r$ should increase linearly with
$a/L_n$. Although, we find the right trend $\omega_r\propto a/L_n$,
{\sc gs2} produces higher values than the slab model. The reason for
this discrepancy is likely that the mode is not purely kink anymore,
but instead develops some kinetic ballooning character due to the finite
pressure gradient drive.

There are small deviations from the $\omega_r=0$ result of the slab
model in the {\sc gs2} simulations in Fig.~\ref{compar}a, b and
d-f. These may be the result of the magnetic drift effects neglected in the
slab model, but also, they may also represent the finite accuracy of the
simulations. In certain cases, when $\sigma$ is very
high, making the parallel mode structure very oscillatory,
exceptionally high parallel resolutions were necessary in {\sc gs2} to
achieve the accuracy presented in Fig.~\ref{compar} (for example 140
grid points in $\theta$).
  
\begin{figure}[htbp]
\begin{center}
\includegraphics[width=0.75\textwidth]{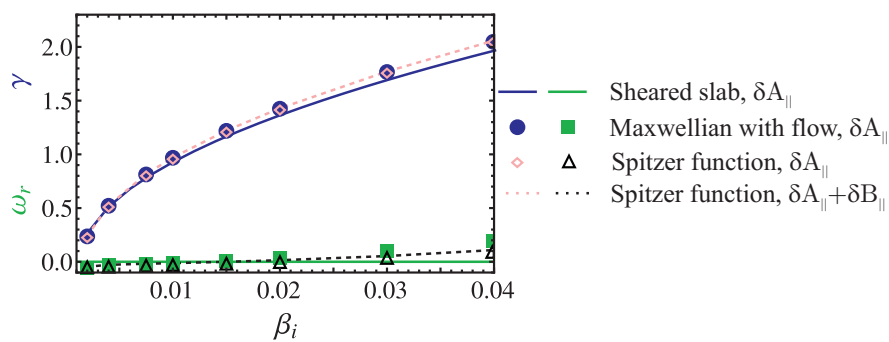}
\end{center} 
\caption{$\beta_i$ scaling of the growth rate $\gamma$ (upper
    curves and points) and real frequency $\omega_r$ of the high mode
    number kink mode (given in $v_i/a$ units). The solid lines are
    sheared slab model results, the dotted curves and symbols are {\sc
      gs2} simulation results computed using a shifted Maxwellian
    electron distribution (full symbols), using a Spitzer function
    keeping only $A_\|$ fluctuations (empty symbols), and using a
    Spitzer function keeping both $A_\|$ and $B_\|$ fluctuations
    (dotted curves). }
\label{betascale}
\end{figure}

 The expected $\sqrt{\beta_i}$ dependence of the growth rate of
  the high mode number kink modes is reproduced, as seen in
  Fig.~\ref{betascale}. Apart from the sheared slab results (solid)
  lines, Fig.~\ref{betascale} shows {\sc gs2} simulations of different
  levels of sophistication. In the simplest case the non-fluctuating
  electron distribution is modeled as a Maxwellian with a finite
  parallel flow velocity (shown with solid symbols). It is interesting
  to see that when the shifted Maxwellian is replaced by a Spitzer
  function with the same flow speed but considerably more complicated
  velocity space structure (given by (B4) and (B8) of \cite{plateau}), the
  results (empty symbols) remain practically unchanged, especially for
  the growth rates. Spot checks for different plasma parameters show
  the same behavior. This demonstrates that the velocity structure of
  the non-fluctuating part of the electron distribution is
  unimportant, and that only its parallel flow speed matters for the
  kink mode. All the simulations presented herein include
  only $A_\|$ perturbations except those shown with the dotted lines in
  Fig.~\ref{betascale}. We find that in the strongly driven cases
  corresponding to our baseline set of parameters, compressional
  magnetic perturbations have no significant impact on the mode
  frequencies.

  We note that the normalized ideal magnetohydrodynamic drive,
  often referred to as the \emph{MHD inertial-layer width}
  \cite{mischenko} is qualitatively different for the high-m kink
  modes studied here and for the $m=1$ mode \cite{rosenbluth}. This
  drive, which determines the ideal MHD growth rate of the mode, is
  $\epsilon^2$ small in the $m=1$ case (as compared to $m\ne 1$)
  making the near marginally stable mode sensitive to non-ideal
  effects such as collisional or collisionless reconnection. Although
  the simulations shown in Figs.~\ref{compar} and \ref{betascale} are
  collisionless and they do not resolve scales of the electron skin
  depth, these high-m modes are so strongly unstable due to the ideal
  MHD drive that they are not expected to be sensitive to physics
  happening in small layers around the $\kpa=0$ surface.

\begin{figure}[htbp]
\begin{center}
\includegraphics[width=1\textwidth]{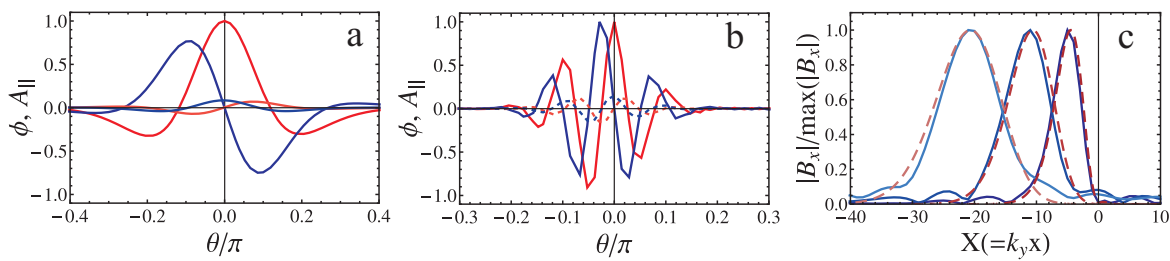}
\end{center} 
\caption{a-b: Parallel mode structures from {\sc gs2}
  simulations. Solid curves are $\phi$, and dashed curves are $\Apa$;
  red and blue curves correspond to the real and imaginary parts,
  respectively; and (a) $\beta_i=0.004$, (b) $\beta_i=0.02$. (c) The
  radial mode structures in the SSM (dashed curves) and calculated
  from {\sc gs2} parallel mode structures (solid curves);
  $\beta_i=\{0.004,\,0.01,\,0.02\}$, the corresponding curves peak at
  increasing $|X|$ values. }
\label{modestruct}
\end{figure}

Typical parallel mode structures are shown in Figures
\ref{modestruct}a and b. These simulations are done for the baseline
parameters with varying plasma beta; $\beta_i=0.004$ and $0.02$ in
\ref{modestruct}a and b, respectively. Note that the kink drive,
$|j_0|'\propto |n_e u|'$, is still finite due to the finite density
gradient.  Increasing $\beta_i$, corresponds to more oscillatory
parallel structures (larger $\kpa$), as expected from
(\ref{optimalkpa}), and an increasing amplitude of the magnetic
component of the fluctuations. In the sheared slab geometry, the
parallel wave number increases away from the resonant surface (recall
$\kpa=k_y x/L_s$).

The Fourier transform of the sheared slab problem in the $x$
coordinate can lead to an equation that is equivalent to the problem
in ballooning representation with a coordinate along the magnetic
field line \cite{balloon}. More precisely, the radial eigenfunction in
the sheared slab, $\hat{B}(X)$, is related to the ballooning
eigenfunction, $B_B(\theta)$, by $\hat{B}(X)\propto
\int_{-\infty}^{\infty} d\theta e^{i\theta X} B_B(-\theta/s)$.  Figure
\ref{modestruct}c shows that the kink modes considered here have this
same property. It compares the variation of the radial mode structure
[the magnitude of $\hat{B}(X)$] in sheared slab calculations (dashed
lines), with the transform of the ballooning mode variation obtained
from {\sc gs2} (solid), for different values of $\beta_i$. The solid
line peaking the closest to (and furthest away from) the rational
surface correspond to the ballooning eigenfunction in
Fig.~\ref{modestruct}a (and b, respectively). The ``ballooning
  character'' of the eigenfunctions, that is, their localization
  around $\theta=0$, is simply a consequence of how a mode with a
  finite radial extent appears in ballooning representation, rather
  than a result of a poloidal dependence in the drive of the mode. In
  particular it is not a magnetic drift effect.  As the radial extent
  of $\Bh(X)$ increases with increasing $\sigma\propto\beta_i$, the equivalent
  $B_B(\theta)$ becomes more and more localized around $\theta=0$
  according to the properties of the Fourier transformation.

We note that from the
sheared slab dispersion relation (\ref{eqjpa03b}) and $\Apa=\kpa
c\pho/\omega$ the long wavelength ballooning equations solved by {\sc
  gs2} can be recovered using the replacements
$i\kpa\rightarrow(qR)^{-1}\p_\theta$ and $i k_y\yh+\xh \p_x\rightarrow
i k_y\yh+ i k_y s\theta\xh$.


\section{Discussion and conclusions}
\label{secconc}
We have developed a procedure for modeling current gradient driven
kink instabilities in a tokamak with {\sc gs2} gyrokinetic simulations
and compared the results to the analytical expressions we derived.

We find that at sufficiently high current gradient high mode number
kink modes are destabilized. The properties of strongly driven kink
modes can be understood from simple analytical expressions derived in a
shearless magnetic geometry by assuming that the mode chooses an
optimal, finite parallel wave number that maximizes its growth
rate. In terms of kinetic quantities, the mode is destabilized by high
$\beta_i$, strong parallel electron flow $u$, high values of
$\p_\psi(\ln n_e u)$, and small perpendicular wave numbers. 

Since the mode is more unstable for smaller values of the
perpendicular wave numbers $\kpe$, magnetic drift effects
($\propto\kv_\perp\cd\vv_{da}$) are unimportant for describing the
stability of the mode. A perhaps more important effect of toroidicity
is that there is a lower limit on $\kpe$ set by the lowest finite
toroidal mode number $n=1$. However, both the analytical calculations
and {\sc gs2} assume a scale separation $\kpa\ll\kpe$ and disregard
global profile and magnetic geometry variations, thus are unable to
properly treat low mode number magnetohydrodynamic modes.  Therefore
the stability limit, which we derive based on kinetic theory,
coincides with the magnetohydrodynamic stability limit for high mode
number kink modes \cite{kadomtsev}. In the sheared slab magnetic
geometry we find that the mode is strongly asymmetric, being localized
on one side with respect to a resonant ($\kpa=0$) surface. The
parallel wave number corresponding to the radial location of the
highest amplitude is close to the one that maximizes the growth rate
in the local theory. The number of unstable radial eigenmodes
increases with increasing drive.

We find good agreement between {\sc gs2} simulations and analytical
estimates both in terms of the parametric dependences of the growth
rates and mode frequencies, and in terms of eigenmode
structure. The large aspect ratio and small $k_y\rho_i$ limit of
  high mode number kink modes may be used as a simple test case for
  linear validation of electromagnetic gyrokinetic codes when current
  drive is to be modeled. By comparing kink modes assuming a
  Maxwellian electron distribution with a parallel flow and
  alternatively a Spitzer function departure from a Maxwellian as a
  drive we demonstrate that the exact velocity structure of the
non-fluctuating electron distribution function is unimportant for the
mode. Only the parallel flow speed of electrons matters.

For modes that are electrostatic in nature, an electron flow -- even
when comparable to the ion thermal speed -- is not expected to
significantly modify their stability. The circulating electrons which
can flow along the field lines are close to be adiabatic, and their
already small non-adiabatic response is only modified by an even
smaller correction from the flow. Without showing specific {\sc gs2}
results, we remark that we have found practically no effect on ion-
and electron temperature gradient modes for typical plasma parameters
even when the plasma $\beta$ and the electron flow speed exceeds their
experimentally relevant range in the simulations.

In a screw-pinch geometry it is known that, if an ideal
magnetohydrodynamic mode is unstable at a given finite poloidal mode
number $m_0$, it should be even more unstable at all mode numbers $m$
satisfying $1\le m<m_0$ \cite{newcomb}. Therefore, the trend of
increasing growth rate with decreasing $k_\perp$ is not terminated
until the lowest wave number allowed in the system. Consequently, if
the plasma is globally stable to low mode number kink modes, it should
be stable for all mode numbers. However, since a similar theorem has
not been proven in toroidal geometry, the relevance of high mode
number kink modes in tokamaks is unclear, and should be the subject of
future investigations. Toward this end, the research herein
demonstrates that suitably modified gyrokinetic codes can be used to
investigate current driven or kink instabilities in
tokamaks. A local code such as {\sc gs2} permits the
  modeling of only high wave number modes, but it has the important
  advantage that it can effectively model the nonlinear evolution of
  these modes, which is a topic for future studies.


\ack The authors are thankful to Jesus Ramos, Jack Connor, Jeff
Freidberg, and Jim Hastie for several fruitful discussions on MHD
related problems, and to Choongki Sung for providing experimental
parameters.  This work was funded by the European Communities under
Association Contract between EURATOM and Vetenskapsr{\aa}det (VR), and
by the US Department of Energy grant at DE-FG02-91ER-54109 at MIT. The
first author is grateful for the financial support of VR.


\appendix
\section{Quasineutrality}
\label{appqn}
To derive the explicit form of the quasineutrality equation from
(\ref{qn2}) we write it as
\begin{equation} 
0=\ea \int d^3v\,\gef-\frac{\ee^2 n_e}{\Te}\pho+\ei \int
d^3v\,\gif-\frac{\ei^2 n_i}{\Ti}\pho,
\label{qnA1}
\end{equation}
where the integrals are taken at fixed particle position. First we
will evaluate the electron contribution to quasineutrality, i.e. the
first two terms of (\ref{qnA1}). We neglect the magnetic drifts in
(\ref{dkel}), replace time derivatives by $-i \omega$, toroidal
derivatives by $-in$, write $\Epa=-i\kpa\pho+i\omega\Apa/c$, and then
divide the equation by $-i\omega+i\kpa\vpa$, to obtain
\begin{eqnarray}
g_e=\frac{\ee}{\Te}\fme\left(1-\frac{\me}{\Te}u\vpa
\right)\frac{\omega\left(\pho-\frac{\vpa}{c}\Apa\right)}{\omega-\kpa\vpa}
\label{geexpr}\\ -nc\fme
\frac{\pho-\frac{\vpa}{c}\Apa}{\omega-\kpa\vpa}\left[\Foe-\frac{\me}{\Te}u\vpa
  \Fte\right]+\fme\frac{\ee}{\Te}u\kpa
\frac{\pho-\frac{\omega\Apa}{\kpa c}}{\omega-\kpa\vpa}.\nonumber
\end{eqnarray}
 The integral $\ee\int d^3v\,\gef$ in (\ref{qnA1}) can be directly
 evaluated in terms of the plasma dispersion function, using that
\begin{equation}
Z(\xi)=\int_{-\infty}^\infty \frac{d
 x}{\sqrt{\pi}}\frac{\exp(-x^2)}{x-\xi},
\end{equation}
where the integration is done along the Landau contour. 
After a straightforward calculation we find that the
electron contribution to the dispersion relation is
\begin{eqnarray}
-\frac{T_e}{\ee^2 n_e}\int d^3v\,\foe \label{elcontr}\\
  =\left(\pho+\bar{A}\right)\left[
  \Bigl(1+\xie\Ze\Bigr)\left(1-\frac{\ose}{\omega}\right)-
  \frac{\ose\ete}{\omega}\left(\xie^2+\Ze (\xie^3-
  \xie/2)\right)\right]\nonumber\\+
\left(\pho+\bar{A}\right)\frac{U\kpa}{|\kpa|}\left\{\Ze-
2\xie\left[\Bigl(1+\xie\Ze\Bigr)\left(1-\frac{\ose}{\omega}
  (1+\eta_u-\ete)\right)\right.\right.
  \nonumber \\ \left.\left. - \frac{\ose\ete}{\omega}\left(\xie^2+\Ze(\xie^3-
  \xie/2)\right) \right] \right\}+\pho\frac{\ose}{\omega}, \nonumber 
\end{eqnarray}
where we introduced $\xia=\omega/(|\kpa| v_a)$, the normalized flow
speed $U=u/v_e$, the diamagnetic frequency $\osa=(ncT_a/e_a)\p_\psi
n_a$, and $\bar{A}=-\omega \Apa/(\kpa c)$.

Once the ion magnetic drifts are neglected, the gyro-averages
$\langle\cd\rangle$ are replaced by $J_0(\kpe\vpe/\Omega_i)$, and the
$\zeta$-derivatives are written in terms of $\osi$, $g_i$ from
(\ref{gkion}) can be easily expressed as the familiar form
\begin{equation}
g_i=\fmi\frac{\ei}{\Ti}\left(\pho-\frac{\vpa}{c}\Apa\right)
J_0\left(\frac{\kpe\vpe}{\Omega_i}\right)
\frac{\omega-\osi\left[1+\eti\left(\frac{\mi
      v_i^2}{2\Ti}-\frac{3}{2}\right)\right]}{\omega-\kpa\vpa}.
\end{equation}
When we evaluate the velocity integral for ions in (\ref{qnA1}) we
expand in the FLR parameter, writing
$J_0^2(\kpe\vpe/\Omega_i)=J_0^2(\kpe\rho_i \vpe/v_i)\approx
1-\alpha_i(\vpe/v_i)^2$, where we recall the definition
$\alpha_i=(\kpe\rho_i)^2/2$. The ion contribution to the dispersion
relation, normalized to $-\ei^2n_i/\Ti$, is obtained to be
\begin{eqnarray}
-\frac{T_i}{\ei^2 n_i}\int d^3v\,\foi \label{ioncontr} \\
=(1-\alpha_i)\left(\pho+\bar{A}\right)\left[\Bigl(1+\xii\Zi\Bigr)
\left(1-\frac{\osi}{\omega}\right)\vphantom{\frac{\xi}{2}}
-\frac{\osi\eti}{\omega} \left(\xii^2+\Zi(\xii^3-\xii/2)\right)
\right]\nonumber\\ +\alpha_i\frac{\osi\eti}{\omega}\left[\left(\pho+\bar{A}\right)\xii\Zi+\bar{A}\right]+\alpha_i\pho\left(1-\frac{\osi}{\omega}\right)+
  \pho\frac{\osi}{\omega} \nonumber.
\end{eqnarray}
Note that (\ref{elcontr}) and (\ref{ioncontr}) contain the
contributions from the adiabatic responses.  When the perturbed
quasineutrality equation (\ref{qnA1}) is formed the contributions from
$\pho\ose/\omega$ and $\pho\osi/\omega$ [the last terms in
  (\ref{elcontr}) and (\ref{ioncontr}), respectively] cancel for a
pure plasma, due to quasineutrality $\ee n_e+\ei n_i=0$ and $(\ln
n_e)'=(\ln n_i)'$.

Due to the high electron thermal speed $\xie$ is typically small. As
long as $\xie$ is not much larger than unity there are
$\mathcal{O}(1)$ terms multiplying $\pho+\bar{A}$ in the electron
contribution to quasineutrality (\ref{elcontr}).  In the ion
contribution (\ref{ioncontr}), the terms in the last line, which
cannot be factorized by $\pho+\bar{A}$ are multiplied by $\alpha_i$
that is assumed to be small in our expressions. In conclusion, the
quasineutrality equation has an order unity part that can be
factorized by $\pho-\omega \Apa/(\kpa c)$, and the rest is small in
$\alpha_i\pho$. This means that to satisfy quasineutrality, either
$\pho$ and $\omega \Apa/(\kpa c)$ should nearly cancel or the
coefficient factorized by $\pho-\omega \Apa/(\kpa c)$ should be close
to zero.

\end{document}